\documentclass[epj,final]{svjour}
\usepackage{graphics,epsfig}

\begin{document}

\title{Error Diagrams and Temporal Correlations in a Fracture Model with
  Characteristic and Power-Law Distributed Avalanches} 

\author{Yamir Moreno\inst{1,2}\thanks{e-mail: yamir@posta.unizar.es}
  \and Miguel V\'azquez-Prada\inst{1} \and Javier B. G\'omez\inst{3} \and Amalio
  F. Pacheco\inst{1,2}}

\institute{Departamento de F\'{\i}sica Te\'orica, Universidad de
Zaragoza, Zaragoza 50009, Spain \and Instituto de Biocomputaci\'on y
F\'{\i}sica de Sistemas Complejos, Universidad de Zaragoza, Zaragoza
50009, Spain \and Departamento de Ciencias de la Tierra, Universidad
de Zaragoza, Zaragoza 50009, Spain}

\date{Received: \today / Revised version: }

\abstract{ Forecasting failure events is one of the most important
problems in fracture mechanics and related sciences. In this paper, we
use the Molchan scheme to investigate the error diagrams in a fracture
model which has the notable advantage of displaying two completely
different regimes according to the heterogeneity of the system. In one
regime, a characteristic event is observed while for the second regime
a power-law spectrum of avalanches is obtained reminiscent of
self-organized criticality. We find that both regimes are different
when predicting large avalanches and that, in the second regime, there
are non-trivial temporal correlations associated to clustering of
large events. Finally, we extend the discussion to seismology, where
both kinds of avalanche size distributions can be seen.}


\titlerunning{Error Diagrams and Temporal Correlations...}

\maketitle

\section{Introduction}
\label{section0}

The fracture of heterogeneous materials has been the subject of
intensive research since many years and is one of the oldest
concerns of science \cite{hans90}. This is not only due to the evident
practical and potential profits, but also because the understanding of
fracture processes at a basic level has shed light on other phenomena
like earthquake occurrence \cite{turcotte,sornette00}. Relatively
simple models such as random resistor networks \cite{arcan85}, beam
networks \cite{roux85}, spring-block models \cite{burridge,ofc1,ofc2}
and the so-called Fibre Bundle Models (FBMs)
\cite{hans90,sornette00,daniels,newman95,hemmer,moreno00} have
guided our way to more complex models of fracture. To date, we
actually know many of the failure properties and its dynamics. Yet,
the range of phenomena associated with fracture has not been cast into
a definite physical and theoretical treatment. Some years ago, there
was a burst of activity in what today we know as self-organized
criticality \cite{bak1,bak2}, a theoretical framework that was soon
applied to the study of fracture processes and avalanche like
phenomena in disordered system. Soon afterwards, some interesting
claimings were put out as, for example, the suggestion that the earth's
crust is in a self-organized critical state, being the
Gutenberg-Richter (GR) law \cite{gr} for the frequency of earthquakes
with a given magnitude a result of this self-organization process
\cite{bak3,sornette89,ito}.

From a more practical point of view, perhaps the most important
concern related to fracture processes that remains unsolved is their
predictability \cite{sornette00,kb02}. One is interested in knowing
not only under what circumstances a material will fail or how the
distribution of energy releases looks like, but also {\em when} a
material will break down or an event of a given magnitude (usually we
are interested in largest quakes) will take place. This problem is far
from being solved and different approaches have been adopted in order
to increase the predictive power of actual methods \cite{kb02}. For
instance, in seismicity, there is an unavoidable degree of
arbitrariness in classifying earthquakes. When a relative big
earthquake occurs, it can be either a foreshock of a larger subsequent
event, an aftershock of a preceding large quake or the mainshock
itself. Thus there is no way to differentiate these events in real
time and many algorithms and methods are useless from a practical
point of view as they perform only a posteriori. However, regardless
of how a big event may be classified, the forecasting of large quakes
is the main goal to be achieved. This is the type of challenge that we
face here.

In this paper, we analyze a dissipative fibre bundle model of fracture
with only one parameter that characterizes the heterogeneity degree of
the system \cite{moreno99}. Noticeably, the model exhibits two very
distinct failure regimes obeying different avalanche size
distributions. This allows the study of the occurrence of large
avalanches, for the two phases, within the same model. In one of these
regimes, the avalanche size distribution shows a power law regime for
small events separated by a gap from the larger events, which are of
the order of the system size. This regime is reminiscent of the
characteristic earthquake behavior found in some seismic faults
\cite{wes94,sieh96}. On the other hand, in the second regime, the
system self-organizes into a critical state characterized by a unique
power law avalanche distribution similar to the GR law. Following the
Molchan method \cite{molchan} we compute the error diagrams that
quantify the forecasting of large cascade events and analyze their
correlations in time. Our results could be potentially applied to the
study of earthquake occurrence.

\section{Failure Model}
\label{section1}

The basic ingredients and main features of the model used here
are as follow \cite{moreno99}:

\begin{enumerate}

\item{A set of $N$ elements is located on the sites of a supporting
  square lattice of linear length $L$. Each element represents a fibre, or in
  general terms, a small volume fraction of the material under study.}
\item{To each element $i$, one assigns a random threshold strength
  $\sigma_{i_{th}}$ taken from a probability distribution. The
  threshold values represent the maximum load each of the elements can
  support before failure occurs. We use henceforth a Weibull
  distribution $P(\sigma_{th})=1-e^{-(\sigma_{th}/\sigma_0)^{\rho}}$ to assign
  the failure thresholds assuming a reference load
  $\sigma_0=1$. $\rho$ is called the Weibull index; the bigger it is,
  the narrower the range of threshold values is \cite{moreno99}.}
\item{At each time step, the load on the system is increased
  by calculating the minimum load required for one
  element to break and adding this amount to all the elements, {\em
  i.e.}, $\sigma_i\rightarrow\sigma_i+\nu$, where
  $\nu=min\{\sigma_{i_{th}}-\sigma_i\}$.}
\item{Once the load acting on an element surpasses its failure
  threshold, it fails and the load it bears is equally redistributed
  among {\em all} the elements of the set {\em irrespective} of their
  state. This amounts to a global fibre bundle version where long
  range interactions are assumed \cite{note1}.}
\item{The fraction of load that would correspond to already failed
  elements is simply dissipated so that the model is
  {\em nonconservative}. Moreover, the degree of dissipation depends on both
  the avalanche size and its internal structure \cite{moreno99}.}
\item{The redistribution of load to non-broken fibres may trigger more
  failure events. In such a case, step $4$ is repeated until the
  system reaches a state where all the elements support a load below
  their respective threshold values.}
\item{When an avalanche has come to an end, the external driving, step
  $3$, is repeated and the elements that have been broken in the
  preceding avalanche are healed assigning to them new random strengths
  and a load equal to zero.}

\end{enumerate}

\begin{figure}[!th]
\centerline{\psfig{file=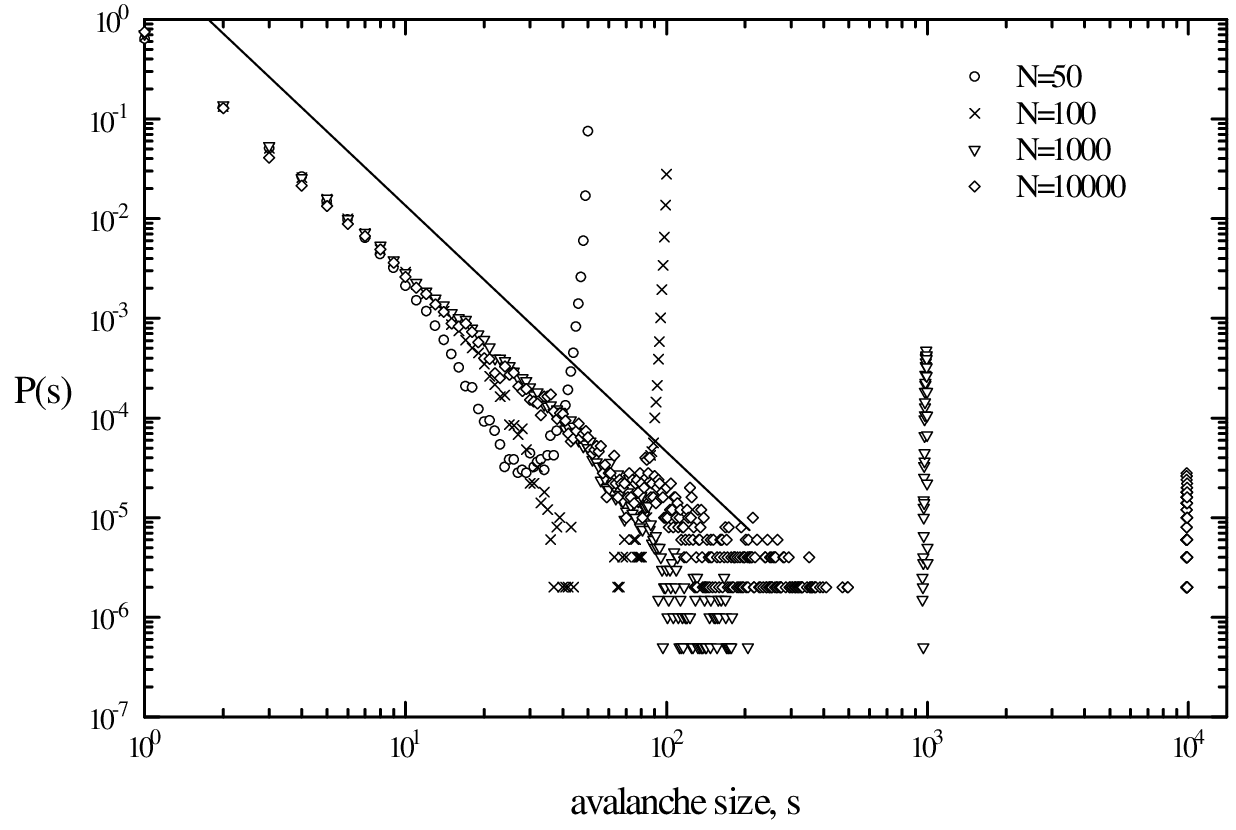,width=2.9in,angle=0}}
\centerline{\psfig{file=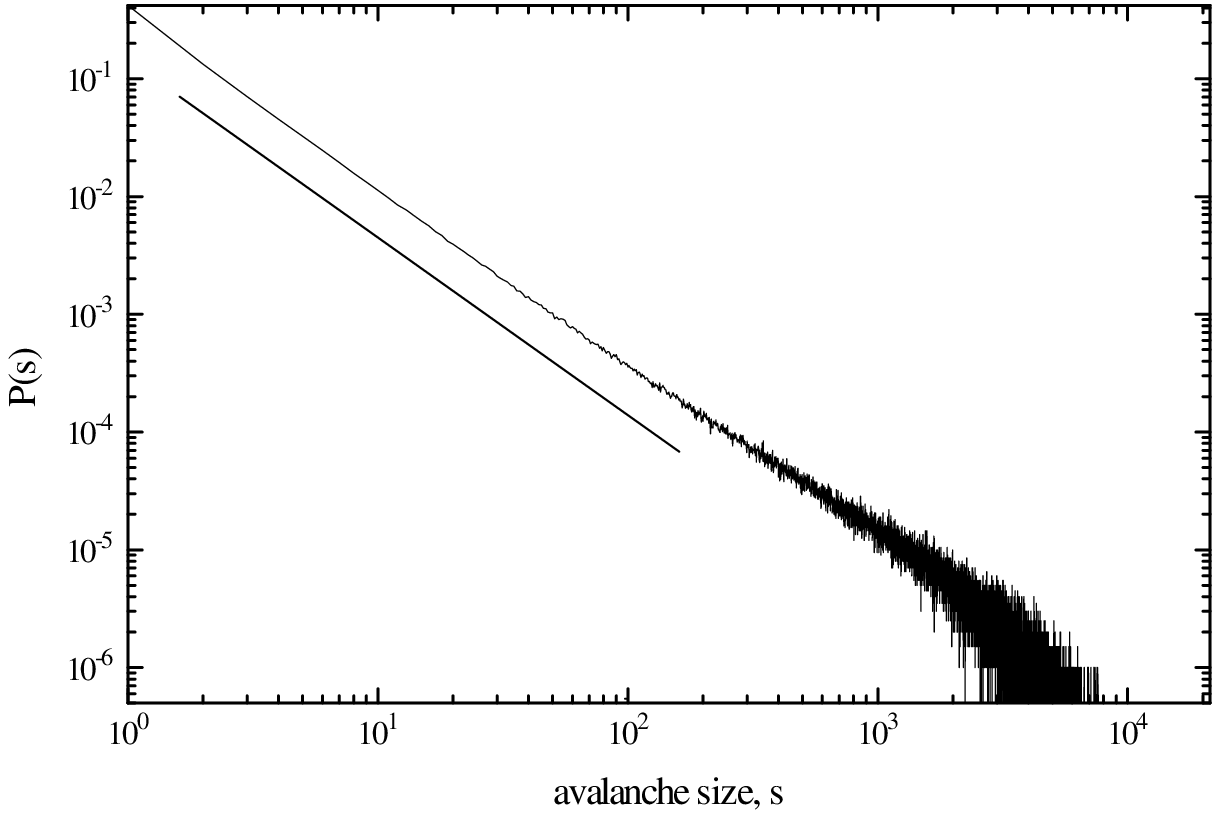,width=2.9in,angle=0}}
\caption{Distributions of avalanche sizes for the two failure regimes
  that characterize the model. The upper part corresponds to a system
  that can be regarded as fairly homogeneous ($\rho=4$) while the lower
  figure is obtained for a more heterogeneous system made up of $N=50000$
  fibres ($\rho=2$). The solid lines have slopes $-2.5$ and $-1.5$
  respectively. After \cite{moreno99}.}
\label{figure1}
\end{figure}

The model is thus constructed by implicitly taking into account some
general properties of any fracture process. Two ingredients of our
model are worth mentioning. First, we assume dissipation through
broken elements. This amounts to consider that the energy dissipated
during a given fracture event is proportional to the size and the
duration of that event. Besides, we incorporate healing only after an
avalanche has ended because when the ongoing avalanche is developing,
the elements have no time for healing and broken regions should not
accumulate stress in that time interval. We additionally note that the
results presented here are valid, strictly speaking, in the limiting
case of very slow driving.

There are two different time scales in the model \cite{moreno99}. One
refers to the external driving of the system and defines the {\em
natural} time interval between failure events. We measure the time
elapsed until a given event $k$ as the sum of the incremental
amounts of load added at every external loading of the system prior to
the event $k$. This definition of time assumes that the rate at which
the system is loaded can be considered constant, and hence that $\nu$
is linearly proportional to the time elapsed since the last
event. This approximation holds, for example, in seismology. The
second time scale of the process quantifies the avalanche lifetimes,
and it is measured as lattice updates. That is, we consider the
avalanches to be instantaneous with respect to the external driving
and separated by an interval of time whose length is proportional to
the amount of energy added to the system during the last external
input.

The iteration of the rules sketched above leads the system to two
different behaviors depending on the level of heterogeneity assumed
when assigning the threshold values \cite{moreno99}. Figure\
\ref{figure1} shows the distributions of avalanche sizes, $P(s)$, for
the two failure regimes of the model. The upper panel shows the
pattern typical of characteristic quakes which are distributed
quasi-periodically in time. This regime, where there are large
fluctuations of the load on the system, is obtained when the bundle of
fibres is homogeneous (large values of the Weibull index $\rho$), that
is, when the deviation of the thresholds from a mean value is not
large. Note that as the system size is increased, large events are
better resolved and there are no intermediate avalanches. 

A richer and more complex behavior is found when the system departs
from homogeneity and the bundle of fibres gets heterogeneous (small
values of $\rho$). In this case, irrespective of the system size, a
power law distribution for the avalanche sizes is obtained. This is
depicted in the lower panel of Figure\ \ref{figure1} for $\rho=2$ and
$N=50000$ fibres.  As we shall see in the next section, the
quasi-periodic nature of the first regime makes the forecasting of
large avalanches feasible and eventually, for large system sizes
(strictly speaking, in the thermodynamic limit), one can always
forecast such events with no errors. This is not anymore the case when
the system can be regarded as heterogeneous, as non-trivial temporal
correlations and clustering of large events come into play making the
forecasting of catastrophic avalanches a hard task.

\section{Error Diagrams and Temporal Correlations}
\label{section2}

\begin{figure}[t]
\centerline{\psfig{file=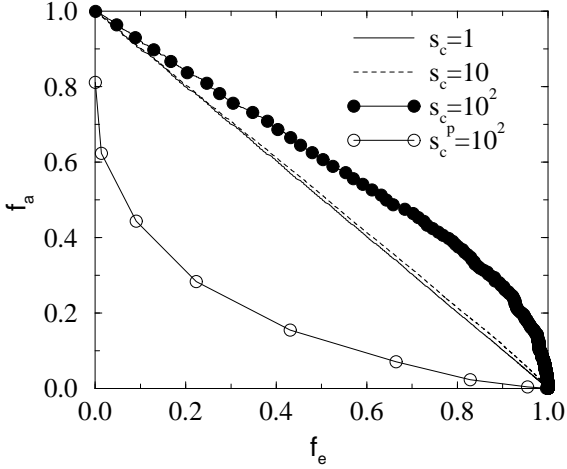,width=2.5in,angle=-90}}
\caption{Error diagrams for the two failure regimes. Open circles
represent the trade-off between $f_a$ and $f_e$ when the system can be
regarded as homogeneous ($\rho=4$ and $N=100$ elements). Full circles
have been obtained for a system made up of $N=10^4$ fibres and
$\rho=2$; the dashed line shows the effects of reducing the target
threshold, $s_c$; the solid line corresponds to the case in which
avalanches of any size are forecasted (equivalent to a random
process).}
\label{figure6}
\end{figure}

In order to inspect whether or not we would be able to predict large
events, we first compute the errors diagrams for both failure
regimes. They are a quantitative measure of the success in forecasting
the occurrence of a given event and were introduced in seismology
several years ago by Molchan \cite{molchan} with the aim of rigorously
evaluating different algorithms for large earthquakes prediction, and
subsequently used by other authors
\cite{kb02,newman02,sornette02,vp03}.

\begin{figure}[t]
\centerline{\psfig{file=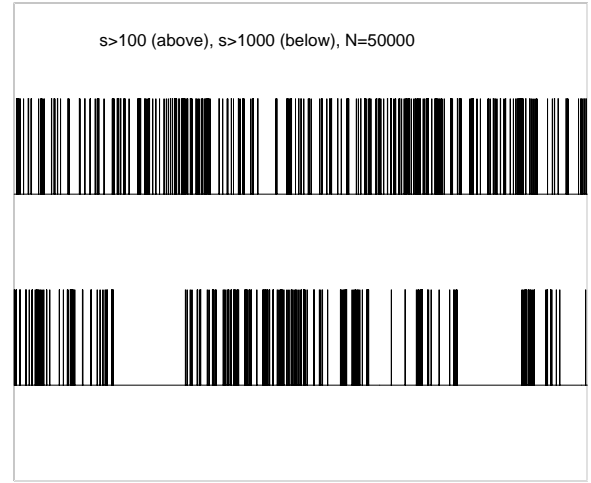,width=2.5in,angle=-90}}
\caption{Time sequence of events, in a model with $\rho=2$, with sizes
equal to or greater than $s$. The time intervals have been rescaled so
that the density of points is the same in both series. The system
consist of $N=50000$ elements. Note the clustering of large quakes
($s>10^3$).}
\label{figure5}
\end{figure}

An error diagram requires the identification of a target to be
forecasted using alarms. Then, one plots the trade-off between the
fraction of target events that were not predicted, $f_e$, and the
fraction of time that the alarm was active, $f_a$. Ideally, one
would like to get the smallest number of failures to predict at the
minimum duration of the alarms. Points close to the origin of an
error diagram represent the best predictions. Hence, the efficiency
of a given forecasting algorithm is measured by how fast $f_e$
decreases when increasing $f_a$. In practice, one switches on the
alarm at a given time interval $\tau$ after the preceding event. If
$\tau=0$, the fraction of failures to predict is zero, but the
fraction of alarm time is 1. On the contrary, if $\tau$ is equal to
(or greater than) the time interval between two successive events,
$f_a=0$ but $f_e=1$. In this way, the result of a completely random
(Poissonian) process is a diagonal line. 

\begin{figure}[th]
\centerline{\psfig{file=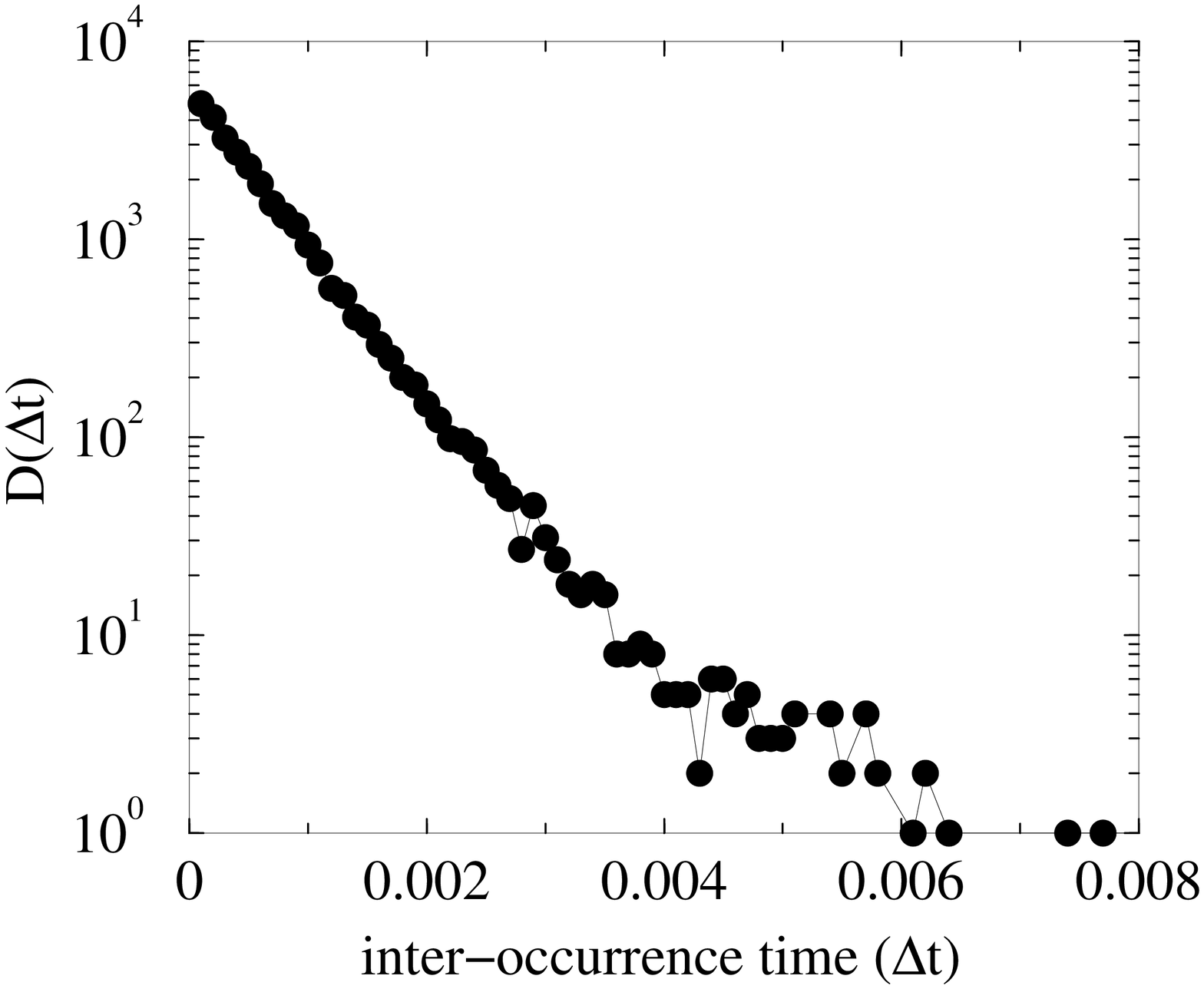,width=2.5in,angle=-90}}
\centerline{\psfig{file=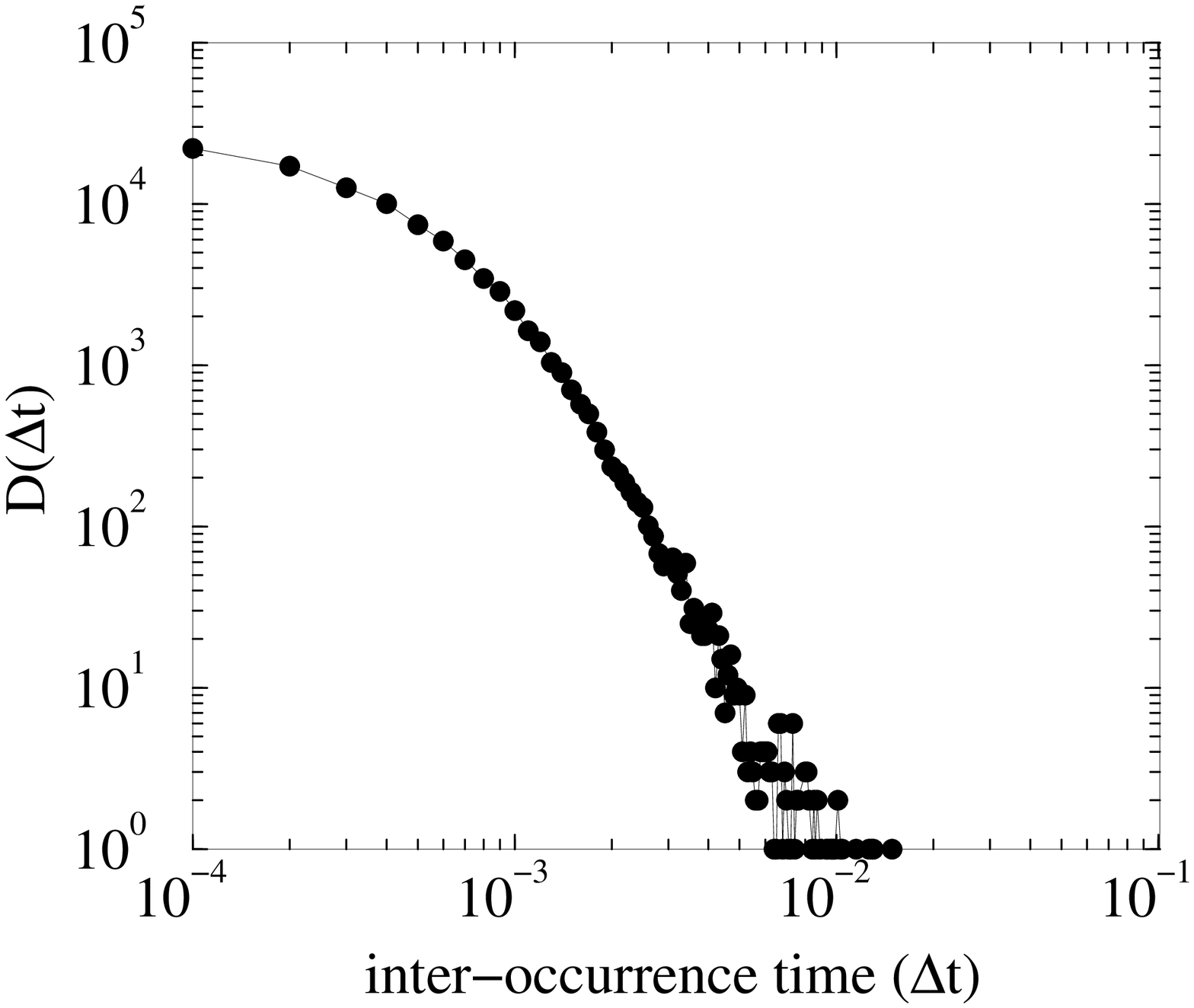,width=2.5in,angle=-90}}
\caption{ Distributions of inter-occurrence time $\Delta t$ for a
system of $N=10^4$ elements and $\rho=2$. The upper figure is in
linear-log scale so that the distribution of $\Delta t$ follows an
exponential law when $s_c$ is small ($s_c>10$). The lower panel
corresponds to $s_c$ equal to or greater than $10^2$ so that only large
avalanches are taken into account. In this case, the inter-occurrence
times are power law distributed.}
\label{figure34}
\end{figure}

Let us now construct the error diagrams for each of the regimes of the
fracture model under study. First, we define the targets as the large
avalanches whose sizes exceed a given threshold $s_c$ and fix the time
interval $\tau$ at which the alarms are switched on after every large
failure event. By varying the length of $\tau$ and computing the
fractions of failures to forecast and of alarm times, $f_e(\tau)$ and
$f_a(\tau)$ respectively, the error diagrams shown in Fig.\
\ref{figure6} are obtained, where the dependency on $\tau$ has been
eliminated. The figure clearly illustrates how different with respect
to successful forecast the two failure regimes are. Open circles
represent the results of the algorithm when applied to the prediction
of characteristic events in the parameter region when the system is
homogeneous. Here, target events occur quasi-periodically and thus
setting $\tau$ close to the mean time interval between large
avalanches ensures a successful prediction with a minimum of alarm
time. For example, one can choose $\tau$ so that $f_a\approx 0.3$ and
$f_e\approx 0.2$, a rather good forecasting. Besides, increasing the
system size improves the prediction of characteristic events since
there appears a gap between small and large events making the
definition of targets less noisy. In this case, the error diagram
approaches a delta function at the origin as large failure events
occur almost periodically in time.

On the contrary, when the system gets heterogeneous, in the error
diagram $f_a$ is always above $f_e$ (full circles). Here, we consider
an avalanche as a target when its size falls on the edge of the
distribution $P(s)$, where finite size effects appear. Reducing the
target thresholds move the error diagram close to the diagonal
line. In fact, the occurrence of an event of any size is a random
process as can be observed in the figure when the threshold is set
equal to 1.

Another interesting information that can be extracted from the error
diagrams for the second regime is the existence of some degree of
non-trivial temporal correlations. The departure from a random process
when the target threshold is increased indicates that large events
show clustering in time. This is clearly appreciated in Fig.\
\ref{figure5}, where two sequences of quakes for a system made up of
$N=50000$ elements and $\rho=2$ has been represented. The upper
series is the occurrence of events with sizes equal to or greater than
$10^2$ while the lower sequence is obtained when $s_c$ is equal to
$10^3$ (upper limit for events not affected by finite-size effects). 

\begin{figure}[t]
\centerline{\psfig{file=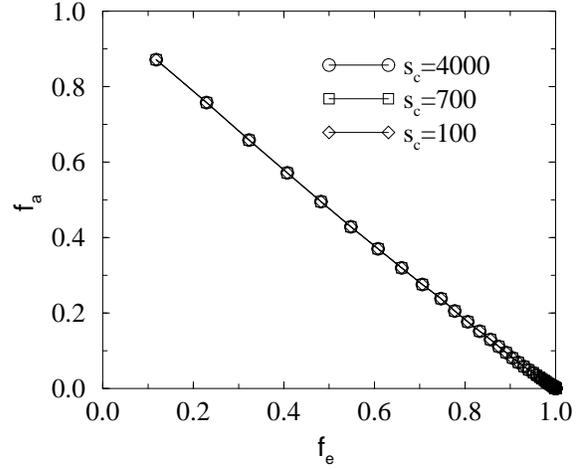,width=2.5in,angle=-90}}
\caption{ Error diagrams of the BTW model for different values of the
target thresholds. Note that irrespective of $s_c$, the forecast is
completely random in agreement with the fact that the BTW model does
not show any kind of temporal correlations. The system size is $N=75
\times 75$.}
\label{fig7}
\end{figure}

We also present in Fig.\ \ref{figure34} the distribution of
inter-occurrence time $\Delta t$ for two values of the target
thresholds. We see that when $s_c$ is increased so that only large
avalanches are taken into account, the distribution $D(\Delta t)$
radically changes from an exponential decay (typical of a random
process) to a decaying power law. This is in agreement with
the error diagrams previously shown and provides further evidence that
small avalanches are not correlated while large ones are. Without
providing a figure, we point out that the distribution of inter-occurrence
time for the first failure regime is a well-behaved function with a clear
mean value and a dispersion that decreases as the system gets larger.

\section{Discussion and Conclusions}

The results shown in the preceding section indicate that the
clustering of large avalanches makes the prediction of such events a
hard task. The shape of the error diagrams and the high rate of
prediction failures can be understood by noticing that the average
inter-occurrence time between {\em all} large events is not a useful
reference in this case. Two other quantities seem to be more
fundamental: the {\em inter-cluster} average time defined as the time
interval between clusters of large avalanches and the mean {\em
intra-cluster} time defined as the time between large events {\em
within} the same cluster. Since the former is much larger that the
latter, the fraction of failures to predict is raised when $\tau$ is
large. On the other hand, setting $\tau$ too small to match the {\em
intra-cluster} time produces a significant increment of $f_a$. Thus
both phenomena contribute to the error diagrams in such a way that the
forecasting is worse than a random prediction.

The clustering of large events is due to the way in which energy is
dissipated and not to the power-law nature of the avalanche size
distribution. The average inter-occurrence time between large quakes
can be considered as a measure of the buildup time for correlations in
the system. When a series of big cascades occurs, the system ends up
in a state where most of the elements are unloaded and a significant
input of energy (or a large time interval) is needed in order to start
another series. In other words, the way in which energy is dissipated
creates stress correlations giving rise on its turn to the observed
clustering. This does not hold anymore when avalanches of smaller
sizes are considered. In this case, the occurrence of a small
avalanche does not significantly alter the state of the system. A
small event can not influence the occurrence of any event at short
times.

The same kind of temporal correlations is seen in other models
displaying self-organized criticality. This is the case of the Olami,
Feder and Christensen (OFC) spring-block model for earthquakes
\cite{ofc1,ofc2}. They found some years ago that non-trivial temporal
correlations of the sort obtained here show up when the model is made
nonconservative. By calculating the coefficient of variation, it was
shown that large earthquakes show clustering while small ones are
uncorrelated in time. Besides, when the model is conservative, neither
spatial nor temporal correlations were found \cite{ofc2}. The fact
that dissipation appears to be a basic ingredient for a model to show
correlations in time and/or space is corroborated in Fig.\ \ref{fig7},
where the error diagrams for the original Bak, Tang and Wiesenfel
(BTW) model \cite{bak1,bak2} are depicted. As can be clearly seen in
the figure, the model can not be distinguished from a random process
with regard to its predictability. Changing the target thresholds does
not improve the forecasting. Thus, in this model, small avalanches and
large ones are completely equivalent and all correlations are ruled
out.

We would like to remark that the results obtained in this paper may be
applied to the study of earthquake occurrence. One of the few
well-known facts in seismology is the clustering of earthquakes in
time and space. While small earthquakes seem to be uncorrelated
\cite{joh85}, large earthquakes display strong clustering
\cite{kagan,sergio}. On the other hand, by tuning a single parameter
in our model, one moves from a characteristic earthquake scenario
\cite{vp02} to a regime where the avalanche sizes are power law
distributed (GR law). The possibility of exploring such different
regimes within the same model is another motivation for further
research in this direction \cite{note2}. 

Summing up, we have analyzed a fracture model with non-trivial
temporal correlations. The results presented seem to support the idea
that models aimed at modeling real earthquakes should incorporate some
kind of dissipation and self-organization. On the other hand, the
clustering of large quakes does not help to improve the predictability
of devastating events since two relevant time scales come into
play. It might be interesting to compute the error diagrams with a
tunable alarm time in such a way that one can better resolve series of
large events and periods of stasis.

\begin{acknowledgement}
Y.\ M.\ acknowledges financial support from the Secretar\'{\i}a de
Estado de Educaci\'on y Universidades (Spain, SB2000-0357). M.\ V-P.\
is supported by the Ph.D grant B037/2001 of the DGA. This work has
been partially supported by the Spanish DGICYT project BFM2002-01798.
\end{acknowledgement}


\begin{thebibliography}{99}

\bibitem{hans90}  H. J. Herrmann, and S. Roux, (Eds), {\em
 Statistical Models for the Fracture of Disordered Media} (North
 Holland, Amsterdam, 1990), and references therein.

\bibitem{turcotte} D. L. Turcotte, {\em Fractals and Chaos in Geology
 and Geophysics}. 2nd ed (Cambridge University Press, Cambridge, New
 York, 1997).

\bibitem{sornette00} D. Sornette,{\em Critical Phenomena in Natural
 Sciences}, (Springer-Verlag, New York, 2000).

\bibitem{arcan85} L. de Arcangelis, S. Redner, and H. J. Herrmann,
 {\em J. Phys. (Paris)} {\bf 46}, L585 (1985).

\bibitem{roux85} S. Roux, and E. Guyon, {\em J. Phys. (France) Lett.}
 {\bf 46}, L999 (1985).

\bibitem{burridge} R. Burridgeq, and L. Knopoff, {\em
Bull. Seismol. Soc. Am.}{\bf 57}, 341 (1967).

\bibitem{ofc1} Z. Olami, H. J. Feder, and K. Christensen, {\em
 Phys. Rev. Lett.} {\bf 68}, 1244 (1992).

\bibitem{ofc2} K. Christensen, and Z. Olami, {\em J. Geophys. Res.} {\bf
 97}, 8729 (1992).

\bibitem{daniels} H. E. Daniels, {\em Proc. R. Soc. London},
  Ser. A{\bf 183},404 (1945).

\bibitem{newman95} W. I. Newman, D. L. Turcotte, and A. M. Gabrielov,
 {\em Phys. Rev. E} {\bf 52}, 4827 (1995).

\bibitem{hemmer} P. C. Hemmer,and A. Hansen, {\em J. Appl. Mech.}
 {\bf 59}, 909 (1992).

\bibitem{moreno00} Y. Moreno, J. B. G\'{o}mez, and A. F. Pacheco, {\em
Phys. Rev. Lett.} {\bf 85}, 2865 (2000).

\bibitem{bak1} P. Bak, C. Tang, and K. Wiesenfeld, {\em
Phys. Rev. Lett.}  {\bf 59}, 381 (1987).

\bibitem{bak2} P. Bak, C. Tang, and K. Wiesenfeld, {\em Phys. Rev. A.}
{\bf 38}, 364 (1988).

\bibitem{gr} G. Gutenberg, and C. F. Richter, {\em Ann. Geophys.} {\bf
9}, 1 (1956).

\bibitem{bak3} P. Bak, and C. Tang, {\em J. Geophys. Res.} {\bf 94},
 15635 (1989).

\bibitem{sornette89} D. Sornette, and A. Sornette, {\em
Europhys. Lett.}  {\bf 9}, 197 (1989).

\bibitem{ito} K. Ito, and M. Matsuzaki, {\em J. Geophys. Res.} {\bf 95},
6853 (1990).

\bibitem{kb02} V. Keilis-Borok, {\em Annu. Rev. Earth Planet. Sci.}
{\bf 30}, 1 (2002).

\bibitem{moreno99} Y. Moreno, J. B. G\'{o}mez, and A. F. Pacheco,
{\em Physica A} {\bf 274}, 400 (1999).

\bibitem{wes94} S. G. Wesnousky, {\em Bull Seismol. Soc. Am.}  {\bf
84}, 1940 (1994).

\bibitem{sieh96} K. Sieh, {\em Proc. Natl. Acad. Sci. USA}  {\bf
93}, 3764 (1996).

\bibitem{molchan} G. M. Molchan, {\em Pure Appl. Geophys.} {\bf 149},
233 (1997).

\bibitem{note1} This choice may seem arbitrary. While the equal load
  sharing is the simplest load transfer rule one may consider, it also
  gives more accurate results when modeling elastic forces than other
  load transfer schemes, at least within the FBM context. However, one
  can in principle consider general load transfer schemes as the one
  introduced in \cite{raul}. There, it is shown that the behavior of
  the system can be regarded as global provided that the interactions
  follows a power law of the form $r^{-\gamma}$ with
  $\gamma\le 2$. On the other hand, similar results have
  been obtained when using local models \cite{ofc}.

\bibitem{raul} R. C. Hidalgo, Y. Moreno, F. Kun, and H. J. Herrmann,
Phys. Rev. E {\bf 65}, 046148 (2002).

\bibitem{ofc} Z. Olami, H. J. Feder, and K. Christensen, Phys. Rev. Lett.
{\bf 68}, 1244 (1992).

\bibitem{newman02} W. I. Newman, and D. L. Turcotte, {\em
Nonlin. Proc. Geophys.} {\bf 9}, 1 (2002).

\bibitem{sornette02} A. Helmstetter, and D. Sornette, {\em e-print}
  cond-mat/0208597 (2002).

\bibitem{vp03} M. V\'{a}zquez-Prada, \'A. Gonz\'alez, J. B. G\'{o}mez,
 and A. F. Pacheco, unpublished (2003).

\bibitem{joh85} A. C. Johnston, and S. J. Nava, {\em J. Geophys. Res.}
 {\bf 90}, 6737 (1985).

\bibitem{kagan} Y. Y. Kagan, and D. D. Jackson, {\em Geophys. J. Int.}
 {\bf 104}, 117 (1991).

\bibitem{sergio} M. S. Mega, P. Allegrini, P. Grigolini, V. Latora,
  L. Palatella, A. Rapisarda, and S. Vinciguerra, {\em
  Phys. Rev. Lett.}, in press. Also {\em e-print} cond-mat/0208597
  (2002).

\bibitem{vp02} M. V\'{a}zquez-Prada, \'A. Gonz\'alez, J. B. G\'{o}mez,
 and A. F. Pacheco, {\em Nonlin. Proc. Geophys.} {\bf 9}, 513 (2002).

\bibitem{note2} Models able to pass from the GR to a
  characteristic earthquake behavior are not found very often in the
  literature. One of such models was proposed in \cite{dahmen98},
  where the concept of configurational entropy is used to argue the
  existence of the two regimes. Our model also suggests that these two
  behaviors can be recognized when the degree of heterogeneity of the
  faults under study are different. 

\bibitem{dahmen98} K. Dahmen, D. Ertas, and Y. Ben-Zion, {\em
Phys. Rev. E} {\bf 58}, 1494 (1998).


\end{thebibliography}
\end{document}